\documentclass[aps,prl,twocolumn,showpacs,groupedaddress,lengthcheck]{revtex4-1}

\pdfoutput=1
\usepackage{amsmath,amssymb,xspace,graphicx,xcolor}
\usepackage[pdftex]{hyperref}
\renewcommand{\r}{{\bf r}}

\newcommand{\blue}[1]{{#1}}
\newcommand{\eps}{\varepsilon}

\begin{document}

\title{Revisiting the flocking transition using active spins}

\author{A. P. Solon$^{1}$, J. Tailleur$^{1}$}
\affiliation{$^{1}$~Univ Paris Diderot, Sorbonne Paris Cite, MSC, UMR 7057 CNRS, F75205 Paris, France}

\date{\today}
 
\begin{abstract}
We consider an active Ising model in which spins both diffuse and
align on lattice in one and two dimensions. The diffusion is biased so
that plus or minus spins hop preferably to the left or to the right,
which generates a flocking transition at low temperature/high
density. We construct a coarse-grained description of the model that
predicts this transition to be a first-order liquid-gas transition in
the temperature-density ensemble, with a critical density sent to
infinity. In this first-order phase transition, the magnetization is
proportional to the liquid fraction and thus varies continuously
throughout the phase diagram. Using microscopic simulations, we show
that this theoretical prediction holds in 2d whereas the fluctuations
alter the transition in 1d, preventing for instance any spontaneous
symmetry breaking.
\end{abstract}


\maketitle

Active matter systems are driven out-of-equilibrium by the injection
of energy \textit{at the single particle
  level}~\cite{Rev1,Rev2,Rev3,Rev4}. This microscopic breakdown of
detailed-balance is responsible for a wide range of phenomena that
have aroused the interest of physicists, from bacterial
ratchets~\cite{Austin,Tailleur2009,diLeonardo,Sokolov} to
self-propelled clusters~\cite{Edinburgh, Lyon, PalacciUS}. \if{}to the
appearance of long-range order in 2d systems with continuous
symmetries~\cite{Vicsek1995,TT}.\fi{} Furthermore, much of this rich
phenomenology is captured by simple models. For instance, the patterns
found in high density motility assays could be accounted for using
simple flocking models~\cite{Bausch, HuguesMT} while clustering in
bacterial mixture was successfully modelled using self-propelled
rods~\cite{PeruaniMB}.

Nevertheless, despite the successful description of many experimental
phenomena, a clear-cut understanding of the underlying mechanisms
sometimes remain elusive. For instance, even though the flocking
transition is one of the central features of active matter, it remains
one of the most debated questions in the field. In their seminal work,
Vicsek and co-workers~\cite{Vicsek1995} showed that self-propelled
particles that align locally can exhibit a transition to long-range
order in 2d. Initially thought to be continuous~\cite{Vicsek1995},
this transition was later shown to be first order using large scale
simulations and a finite-size-scaling akin to that of magnetic phase
transitions~\cite{Gregoire2004}. Many works were also devoted to
nematic~\cite{Peruani2006,cristinaPA,Ginelli2010} or metric-free
interactions~\cite{Ballerni2008}, the latter yielding a
continuous transition~\cite{huguesMetric}. Related flocking models
were also studied in 1d~\cite{Vicsek1d,Evans1999}, where the transition
was found to be continuous, casting even more confusion in the
field. Beyond the presence of strong finite-size effects, a major
difficulty in obtaining conclusive numerical evidence comes from the
lack of a theoretical framework to analyse the finite-size scalings of
flocking models.

In parallel to the numerical studies, much effort was thus devoted to
the construction of such an analytical description of the flocking
transition. While the Vicsek model (VM) is among the simplest to
simulate, it is one of the hardest to coarse-grain, being defined
off-lattice, in discrete time and involving many-body
interactions. Many approaches were thus either
phenomenological~\cite{TT,Mishra2010,Gopinath2012} or focused on
simpler models~\cite{BDG}, and progress is slower for the
VM~\cite{Ihle2011}. Lots of effort was again devoted to the nematic
case~\cite{Baskaran2008,Peshkov2012,Ngo2012} or to topologic
interactions~\cite{Peshkov2012,Ihle2012}. The existence of long-range
order in 2d for polar alignment was established~\cite{TT} but progress
is difficult since the coarse-grained equations are hard to solve
analytically. \blue{Most analytical studies were thus restricted to
  the linear stability analysis of homogeneous solutions or the
  simulation of coarse-grained
  equations~\cite{BDG,Mishra2010,Gopinath2012}. While non-linear
  profiles for a model with \textit{nematic alignment} could be
  computed analytically~\cite{Peshkov2012}, closed analytical
  solutions are still missing for \textit{polar models} despite recent
  progress~\cite{BDG,Mishra2010,Ihle2011}. All in all, in spite of the
  important progresses made during the last few years, a unifying
  theoretical framework of the flocking transition is still missing.}

We present below a tentative step in this direction through the
introduction of a microscopic lattice model with discrete symmetry,
which is much simpler both to simulate and describe analytically than
traditional flocking models. By bridging micro and macro, we show
that the phase diagram of the flocking transition of our model amounts
to a standard liquid-gas transition in the canonical ensemble
\textit{with a critical density $\rho_c=\infty$}. In particular, this
sheds new light on the finite-size scaling of the transition and
predicts the order parameter to vary \textit{continuously} in the
temperature-density plane, in the thermodynamic limit. Furthermore, we
show that there is no continuous transition in 1d, where fluctuations
strongly alter the transition. \if{}Last, the comparison between our
model and the existing literature allows us to disentangle separate
properties observed in flocking models. For instance, there are no
giant number fluctuations in the homogeneous ordered phase of our
model\if{---probably due to the discrete symmetry of the order
  parameter---}\fi{} which shows that such fluctuations are not
intrinsic to polar flocking states.\fi{}

\begin{figure}
  \begin{center}
    \includegraphics[scale=1]{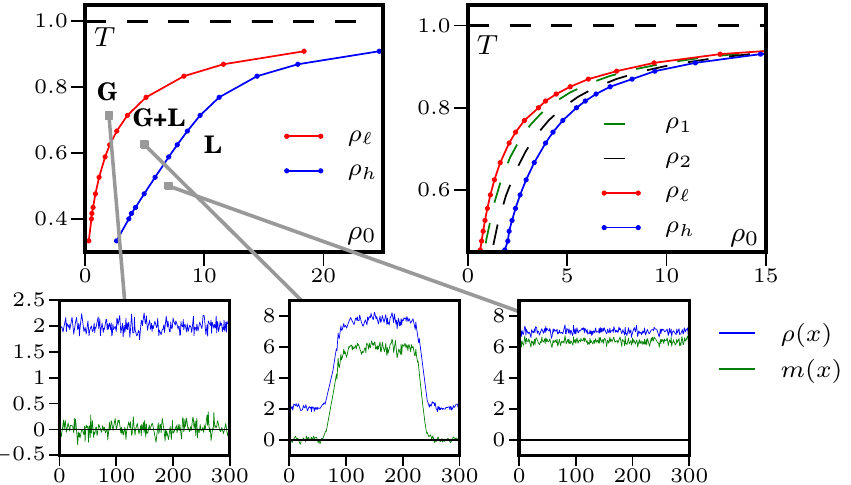}
    \caption{{\bf Top-left}: phase diagram in 2d with ordered liquid
      (L), disordered gas (G), and coexistence region (G+L). The red
      and blue lines correspond to low and high densities of phase
      separated profiles; they enclose the region where such profiles
      can be seen. $D=1$, $\epsilon=0.9$, $L=300$, $\rho_0=N/L$. {\bf
        Bottom}: Snapshots of the different profiles averaged over the
      transverse direction. {\bf Top-right}: Phase diagram predicted
      by the RMFM. In addition to $\rho_h$ and $\rho_\ell$, black and
      green dashed spinodal lines signal the loss of linear stability
      of the homogeneous profiles. $D=v=r=1$.}
    \label{fig:PhaseDiag}
  \end{center}
\end{figure}

Let us consider a one-dimensional lattice of $L$ sites on which $N$
particles have Ising spins $s=\pm 1$. Each particle hops
asymmetrically at rate $D(1+s \eps)$ and $D(1-s\eps)$ to its right and
left neighboring site. (In higher dimensions, the hoping rates are
chosen symmetric in all directions but one.) There is no exclusion
between particles and we note $n_i^\pm$ the numbers of $\pm$ spins on
site $i$ so that the local densities and magnetizations are given by
$\rho_i=n^+_i+n^-_i$ and $m_i=n^+_i-n_i^-$. The particles also align
their spins: on site $i$ a spin $s$ changes sign with rate $\exp(-s
\beta\frac{m_i}{\rho_i})$ where $\beta\equiv 1/T$ plays the role of an
inverse temperature~\cite{footx}.  When $D=0$, the system thus amounts
to $L^d$ independent mean-field Ising models. When $D>0$ and
$\epsilon\neq 0$, three different configurations are typically found
in the system (see Fig.~\ref{fig:PhaseDiag}): at low temperature a
uniform ordered phase, at high temperature a uniform disordered phase,
and in between a phase-separated system with high density ordered
bands ($\rho_i\simeq \rho_h$, $m_i\simeq m_h\neq0$) connected through
narrow interfaces to a disordered homogeneous background
($\rho_i\simeq \rho_\ell$, $m_i\simeq0$). The stability of these
profiles in the thermodynamic limit depends on the number of spatial
dimensions but they are all long-lived in finite systems. Let us now
show how a simple theoretical framework can be constructed to account
for the phase diagram of Fig.~\ref{fig:PhaseDiag}.

Many coarse-graining approaches used in the past rely on factorization
approximation of microscopic kinetic
equations~\cite{Tsimring,BDG,Baskaran2008,Farell}. On a 1d lattice,
this amounts to a simple mean-field approximation: $f(\langle n_i^\pm
\rangle)=\langle f( n_i^\pm) \rangle$, which may be quantitatively
wrong but often captures phase diagrams of lattice-gas models
exactly~\cite{EvansReview} even in complicated
cases~\cite{Sugden}. Introducing continuous variables $x=i/L$, $ v=2 D
\eps/L$ and $\tilde D=D/L^2$, the mean-field dynamics of the
coarse-grained fields $\rho(x)=\langle \rho_i\rangle$ and
$m(x)=\langle m_i \rangle$ is given, in the large $L$ limit, by
\begin{align}
\dot \rho &= \tilde D \partial_{xx} \rho - v \partial_x m \label{eqn:MFrho}\\
\dot m &= \tilde D\partial_{xx} m - v\partial_x \rho + 2 \rho \sinh\frac{\beta m}{\rho} -2m \cosh\frac{\beta m}{\rho}\label{eqn:MFm}
\end{align}
In higher dimensions, one simply replaces $\partial_{xx}$ by a
Laplacian $\Delta$ and we use this more general form hereafter.

Looking for the onset of a flocking transition, we
linearize the dynamics for $m \ll \rho$, which yields~\cite{foot}
\begin{equation}
  \dot m = \tilde D\Delta m -v \partial_x \rho + 2 m (\beta-1) -\alpha \frac{m^3}{\rho^2}\label{eqn:MFmlin}
\end{equation}
where $\alpha=\beta^2(1-\frac \beta 3)$. The profile
$\rho(x)=\rho_0,\,m(x)=0$ is thus linearly unstable for $\beta>1$,
where simulations of Eqs.~\eqref{eqn:MFrho}-\eqref{eqn:MFm} show that
clusters are never stable~\cite{foot1} and always spread to reach the
homogeneous ordered profile $m(x)=m_0$. Mean-field thus predicts a
continuous transition from $m\equiv\frac 1 L \sum_i m_i =0$ to
$m=m_0(\beta)$ at $\beta_c=1$, in clear contradictions with
Fig.~\ref{fig:PhaseDiag}. As often~\cite{Tailleur2008,Thompson2011},
this approximation is only valid for $\rho\to\infty$; for finite
densities we thus expand the mean-field critical temperature to include $1/\rho$
corrections~\cite{foot2} and use $\beta_c\equiv 1+\frac{r}{\rho}$ in
Eq.~\ref{eqn:MFmlin}:
\begin{equation}
    \dot m = \tilde D\Delta m -v\partial_x \rho + 2 m (\beta-1-\frac r \rho) - \alpha \frac{m^3}{\rho^2}\label{eqn:MFR}
\end{equation}

The phase diagram corresponding to Eqs.~\eqref{eqn:MFrho}
and~\eqref{eqn:MFR}, which form our refined mean-field model (RMFM),
is presented in the top-right corner of Fig.~\ref{fig:PhaseDiag}. When
$T<1$, homogeneous disordered (resp. ordered) profiles are always
linearly stable at low enough density $\rho_0<\rho_1$ (resp. high
enough density $\rho_0>\rho_2$). Since $\rho_1<\rho_2$, there is a
finite intermediate region $[\rho_1,\rho_2]$ where neither homogeneous
profiles are stable. In this region, the system separates in two
homogeneous phases connected with sharp fronts: a disordered region
with low density $\rho_\ell<\rho_1$ and an ordered region with high
density $\rho_h>\rho_2$ and $m_h\neq 0$.

Propagating shocks can be computed analytically when $\beta$
is close to 1 by linearizing Eq.~\eqref{eqn:MFR} around the density
$\rho_1=r/(\beta-1)$ at which the homogeneous disordered profile
becomes linearly unstable. We first solve Eq.~\eqref{eqn:MFrho}, by
neglecting the diffusion term in a reference frame moving at speed
$c$, to get $\rho$ as a function of $m$:
\begin{equation}
\rho(\r)=\rho_\ell+\frac{v}{c}m(\r)\label{eqn:rhoofm}
\end{equation}
Eqs.~\eqref{eqn:MFR} and~\eqref{eqn:rhoofm} then yields for $m$
\begin{equation}
  \tilde D\Delta m+c(1-\frac{v^2}{c^2})\partial_x m+\mu[\rho_l-\rho_1+\frac{v}{c}m(x)]m-\alpha \frac{m^3}{\rho_1^2}=0
\end{equation}
where $\mu=2r/\rho_1^2$. Looking for ascending ($q^+>0$) and
descending front ($q^-<0$) solutions
\begin{equation}\label{eqn:msol}
  m(\r )=\frac{m_h}{2}[1+\tanh(q^{\pm} x)]
\end{equation}
one gets 
\begin{equation}\label{eqn:param}
  c=v\quad q^{\pm}=\pm \frac{m_h \sqrt{\alpha}}{\sqrt{8 \tilde D} \rho_1}\quad m_h=\frac{4 r}{3 \alpha}\quad \rho_\ell=\rho_1-\frac{4 r}{9\alpha}
\end{equation}
Such solutions are consistent with our approximations since
$\frac{\rho-\rho_1}{\rho_1}\ll 1$ and $\tilde D \Delta \rho \ll v
\partial_x \rho$ when $\beta\to 1$~\cite{SI}. In this regime, simulations of the
RMFM and Eqs.~(\ref{eqn:rhoofm}-\ref{eqn:param}) yield the same
profiles and band velocities. For larger $\beta$, the $\tilde D
\Delta \rho$ term makes front and rear interfaces asymmetric and $c>v$: the
flocks fly faster than the birds~\cite{SI}.

Since $\rho_\ell$, $\rho_h$ and $m_h$ do not depend on $\rho_0$,
increasing the average density at fixed temperature simply increases
the fraction of the high-density region. In the thermodynamic limit,
phase separated profiles can be seen from $\rho_\ell$ to $\rho_h$. One
always has $\rho_\ell<\rho_1<\rho_2<\rho_h$ so that the clusters and
the homogeneous profiles are both linearly stable in the intervals
$[\rho_\ell,\rho_1]$ and $[\rho_2,\rho_h]$.

\begin{figure}
  \centering
  \includegraphics[scale=.9]{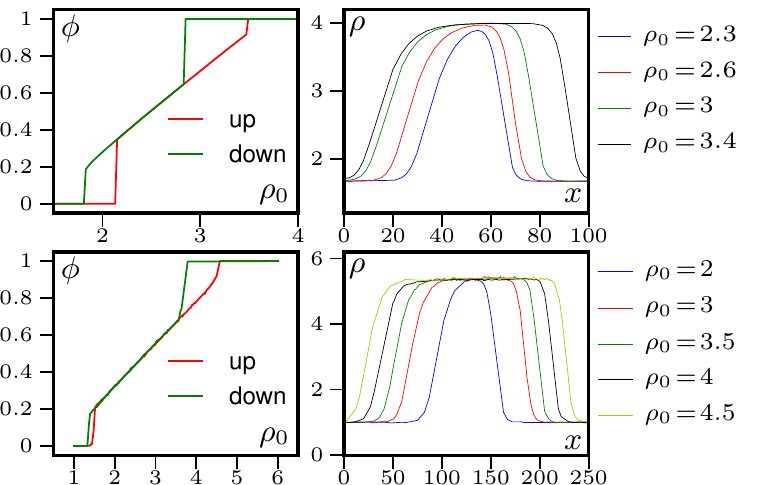}
  \caption{{\bf Left}: Fraction of the ordered liquid phase when
     $\rho_0$ is either increased or decreased for the RFMF
    (top) and in 2d microscopic simulations (bottom). {\bf Right}
    Corresponding profiles of the system. {\bf Parameters}: RMFM
    $L=100$, $v=D=1$, $r=1.6$, $\beta=1.75$, $\Delta \rho_0=10^{-2}$
    every $\Delta t=15\,000$; 2d lattice model $L=250$, $\beta=2$,
    $D=1$, $\eps=0.9$, $\Delta \rho_0=10^{-2}$ every $\Delta t=500$.}
  \label{fig:hysteresis_MF}
\end{figure}

The refined mean-field scenario thus resembles an equilibrium
liquid-gas phase transition in the temperature-density ensemble, the
total magnetization being proportional to the fraction of the liquid
phase. Varying the density $\rho_0$ at fixed $T$, one indeed observes
the traditional hysteresis loops shown in
Fig.~\ref{fig:hysteresis_MF}. Increasing $\rho_0$, homogeneous
disordered profiles are seen up to $\rho_1$ where a discontinuous jump
takes the system into a phase-separated profile. A further density
increase results in a widening of the liquid phase which almost fills
the system when $\rho\lessapprox\rho_h$. (The finite widths of the
interfaces connecting $\rho_\ell$ and $\rho_h$ prevent phase-separated
profiles for $\rho_0$ close to $\rho_{\ell/h}$ in finite
systems.) Going down, the homogeneous liquid phase remains metastable
until $\rho_0=\rho_2$ and discontinuously jumps to a coexistent
state. The fraction of gas then increases until it fills the system at
$\rho\simeq \rho_l$.

Unlike equilibrium liquid-gas transitions, dense and dilute phases in
flocking models have different symmetries due to the coupling between
density and orientation. One thus cannot circumvent the transition and
continuously transform the system from a gas to a liquid: the
transition line cannot stop at a finite point in the $(T,\rho_0)$
plane and, indeed, the critical density is infinite. As far as we are
aware, this has not been described for other flocking
models~\cite{foot3} even though it should be generic and is consistent
with published numerical results on the VM~\cite{Gregoire2004,BDG}.

Let us now turn to simulations of the 2d active Ising model. Beyond
the structure of the phase diagram (see Fig.~\ref{fig:PhaseDiag}), the
RMFM correctly captures the mechanism of the transition. The
coexistence between homogeneous and phase-separated profiles is
confirmed and changing $\rho_0$ at fixed $\beta$ inside the
coexistence region simply changes the fraction of the liquid phase
(see Fig.~\ref{fig:hysteresis_MF} and~\cite{SI}); the velocity of the
high density bands, for instance, remains constant~\cite{SI}. Since
the high density bands have a minimal size $\ell_c$, the apparition of
a flock in a finite-size system corresponds to a discontinuous jump to
a non-zero magnetization $m_0\simeq m_h \ell_c /L$ which vanishes as
$L\to\infty$. As expected for a liquid-gas transition, the order
parameter thus varies \textit{continuously} thoughout the phase
diagram of the canonical ensemble, in the thermodynamic limit.

\begin{figure}
  \includegraphics{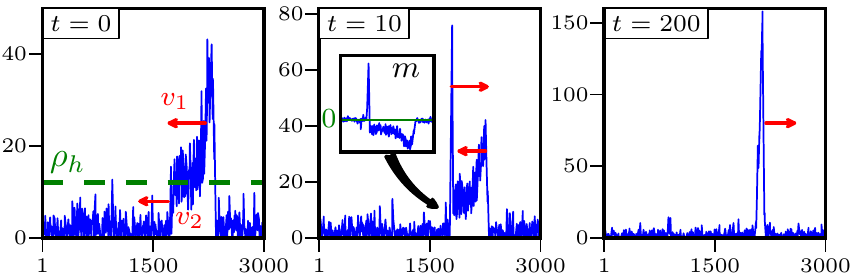}
  \caption{Reversal of a 1d cluster due to a localized
    fluctuation. $v_2$ is greater than $v_1$ until
    $\rho(x)\!=\!\rho_h$ in the whole cluster. (See movies in~\cite{SI}.) $\rho_0=5$, $D=1$,
    $\eps=0.9$, $\beta=1.7$.}
  \label{fig:flips}
\end{figure}

The scenario presented here can be related to the measurement of the
binder cumulant $G=1-\frac{\langle m^4\rangle}{3 \langle m^2 \rangle
  ^2}$ done in the literature~\cite{Gregoire2004,Ngo2012}. The
coexistence between phase-separated profiles and supercooled gas phase
yields a three-peak structure for $P(m)$ around $m=0$ and $m=\pm m_0$
whose relative weights vary across the transition. (The same holds for
the coexistence with superheated liquid.)  Assuming a sum of three
Gaussians of variance $\sigma$, the minimum of $G$,
$G_{\textsuperscript{min}}\!\!=\!\!-[12(\sigma/m_0)^2+36(\sigma/m_0)^4]^{-1}$, is only markedly
negative when $m_0\gg \sigma$. Contrary to  what happens in a
grand-canonical ensemble, \textit{both} $m_0$ and $\sigma$ vanish when
$L\to \infty$, the negative peak does not necessarily becomes more
pronounced as $L\to\infty$, and one can easily mistake a first-order
transition for a 2nd order one if $\sigma$ remains comparable to $m_0$
(see the 1d case below). \if{}Last, many studies use noise rather
than density as a control parameter, which makes the finite-size
scaling even more difficult since $\rho_\ell$, $\rho_h$, $m_h$---and
whence $m_0$---also depends on $T$.\fi{}

Let us now show that fluctuations strongly alter the transition in
1d. First, all three profiles shown on Fig.~\ref{fig:PhaseDiag} exist
and are linearly stable in finite systems~\cite{foot4}. The general
scenario predicted by the RMFM thus holds: homogeneous profiles
between $\rho_1(T)$ and $\rho_2(T)$ are linearly unstable and tend to
phase-separate between linearly stable low-density disordered regions and
high-density ordered regions.

To assess the impact of fluctuations, let us consider the flipping of
an ordered domain in the coexistence region. In 1d, an excess of, say,
positive spins on \textit{a single site} suffices to flip an
approaching negative cluster (see Fig.~\ref{fig:flips}); this happens
frequently and the total magnetization keeps flipping in this
region. The 2d counterpart of such a fluctuation is an excess of
positive spins on \textit{a transverse band of $\sim L$ sites} in
front of the approaching cluster, which has a negligible probability
when $L\to\infty$. Similarly, the $m=m_0$ homogeneous profile is
unstable in the thermodynamic limit in 1d, which may be why it has not
been observed before~\cite{foot4}. Indeed, even though a fluctuation
that would create a small negative cluster in a uniform profile
$m=m_0>0$ is rare, its probability does not decay exponentially fast
with $L$ since only a finite number of sites have to be flipped. When
$L$ increases, so does the entropy of such localized perturbations,
and the time it takes to exit the homogeneous state thus vanishes when
$L\to\infty$.

\begin{figure}
  \centering
  \includegraphics[scale=1]{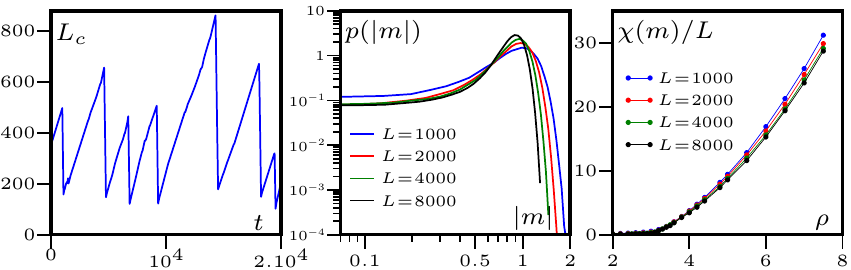}
  \caption{{\bf Left}: Cluster length as a function of time, showing a
    linear spreading between reversals. $D=1$ $\eps=0.9$ $\beta=2$
    $\rho_0=3$. {\bf Center \& Right}: $P(|m|)$ for $\rho_0=4$;
    $\chi_m(\rho)/L$; $\beta=1.538$, $D=1$, $\eps=0.9$.}
  \label{fig:chi}
\end{figure}

In 1d, there are thus only two phases in the thermodynamic limit:
homogeneous disordered profiles and constantly flipping clusters of
opposite magnetization, whose dynamics we now describe. Starting from
a localized cluster, the ordered region spreads at constant speed
(Fig.~\ref{fig:chi}): the fore front is initially faster than the rear
front and their velocity becomes equal only when the density in the
band is uniformly equal to $\rho_h$ (Fig.~\ref{fig:flips} and movies
in~\cite{SI}). The mean cluster size before a reversal, $L_c^R$, is
thus proportional to the mean time between reversals, $\tau_R(L)$. The
reversal then starts when a fluctuation at the front of the cluster
begins to progressively flip all its sites (Fig.~\ref{fig:flips}). It
ends when this fluctuation has travelled through the whole cluster;
this takes a time proportional to $L_c^R$ and thus to $\tau_R(L)$. In
the large size limit, the ratio of the probability of finding the
system in a cluster or in a reversal is thus constant since both the
times spent between and during reversals are $\propto
\tau_R(L)$. $P(m)$ thus has a non-vanishing flat part between $\pm
m_0$ and $\langle m \rangle=0$ (Fig.~\ref{fig:chi}): there is no
spontaneous symmetry breaking in 1d. Consequently, the susceptibility
$\chi_m=L (\langle m^2\rangle-\langle m \rangle ^2)$ is extensive in
the cluster region, as can be seen in Fig.~\ref{fig:chi}. Note that
since the reversals capture a finite part of the steady-state measure,
one should probably not use $|m|$ instead of $m$ when computing
$\chi_m$, as is frequently done for the Ising model and was done in
earlier studies of 1d flocking models.

\begin{figure}
  \includegraphics[scale=1]{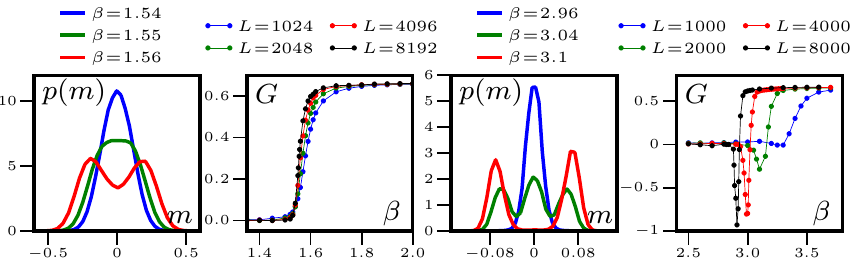}
  \caption{Histograms and Binder cumulant of the total magnetization
    for $\rho_0=3$, $D=1$ (left) and $\rho_0=0.2$, $D=10$
    (right). $\eps=0.9$. $L=8000$ for $P(m)$.}
  \label{fig:contvsdisc}
\end{figure}

The difficulty of analyzing Binder cumulants can be clearly seen in
1d, where the large $L$ limit is easily reached and the three peaks in
$P(m)$ at the transition can be very difficult to
discriminate. If the width of the peaks $\sigma$ is larger than their
separation $m_0$, no negative peak in $G$ is observed and increasing
$L$ does not help since the peaks get closer as they get narrower. In
figure~\ref{fig:contvsdisc} we show two extreme cases: without the
RMFM to analyse the data, it would be very difficult to realize that
one is looking at the same transition. This may explain why previous
studies of 1d flocking models with similar---though not identical---dynamics concluded to a second-order
transition~\cite{Vicsek1d,Evans1999}.

\paragraph{Conclusion}
In this letter we have introduced a lattice model of self-propelled
Ising spins, whose phenomenology is very similar to that of
traditional models of aligning self-propelled particles. The
simplicity of our model allows us to show that its flocking transition
amounts to a liquid-gas phase transition in the canonical ensemble
with an infinite critical density. The total magnetization is
proportional to the liquid fraction and thus varies
\textit{continuously} through this first-order phase transition in the
thermodynamic limit, a rather counter-intuitive result. This scenario,
confirmed by numerical simulations in 2d, is altered by the strength
of fluctuations in 1d, where neither spontaneous symmetry breaking nor
continuous transitions are observed.

\blue{Despite fundamental differences between our model and others
  found in the literature, such as the symmetry of the order
  parameter, many features of the flocking transition observed here
  seem consistent with existing numerical results obtained either for
  microscopic models~\cite{Gregoire2004,Ginelli2010,Peshkov2012} or
  continuous descriptions~\cite{BDG,Mishra2010,Ihle2011,Gopinath2012}
  of self-propelled particles. For instance, the phase diagram seems
  compatible with those of nematic~\cite{Ginelli2010,Peshkov2012} or
  Vicsek models~\cite{Gregoire2004,BDG}, even though the high density
  regions have not been studied in these models. This suggests that the
  analogy between the flocking transition and a canonical liquid gas
  transition could be generic, while the symmetry of the order
  parameter would mostly control features of the ordered liquid
  phase.} For instance, giant-number fluctuations, which have been
reported in flocking models, are trivially present \textit{in the
  coexistence region} of our model. There, $P(\rho_i)$ is
double-peaked (around $\rho_\ell$ and $\rho_h$) and the variance of
the number of particles in a box of finite size satisfies $\langle N^2
\rangle -\langle N \rangle ^2 \propto \langle N\rangle
^2$~\cite{Aranson2008}. They are however absent from \textit{the
  homogeneous ordered phase}~\cite{SI}, which shows that such
fluctuations are not intrinsic to polar flocking states.

\blue{The introduction of Active Spin models is clearly aimed at improving
our theoretical understanding of the flocking transition rather than
accounting for given experiments. One can nevertheless wonder whether
such models could be relevant for experimental systems. Discrete
symmetry of the order parameter can for instance stem out of a
geometry which imposes only two possible flocking directions, as is
the case for locusts constrained in a ring-shaped
arena~\cite{Couzin2006}. Then, as for the Vicsek model, the high
density region can only be attained if the interparticle interaction
range is much larger than the particle size as can be the case, for
instance, for electrostatic, hydrodynamic or social interactions. In
other cases, such as hard rods, one cannot neglect the steric
exclusion between particles and other density-induced effects which
can strongly alter the flocking transition~\cite{Farell}. More
generally, recent progresses on the manipulation of cold atoms in
optical lattices have given a large freedom to control the
interactions in spin chains~\cite{Fukuhara2013}. This could provide
an interesting path to build the quantum version of more general
active spin models.}

Last, part of the difficulty of analysing the transition comes from
the ``choice'' of the temperature-density ensemble where no
discontinuous jump of the magnetization is seen in the thermodynamic
limit. The design of grand-canonical counterparts of flocking models,
in which the magnetization would jump discontinuously at a transition
line, thus seems a promising line of research, even though ``changing
ensemble'' is not obvious out-of-equilibrium.

\begin{acknowledgments}
  The authors thank M. Cheneau, H. Chat\'e, G. Gr\'egoire,
  P. Krapivksy, F. Peruani, H. Touchette, F. van Wijland for useful
  discussions.
\end{acknowledgments}


\begin{thebibliography}{99}
\bibitem{foot} For $\beta>3$ we would need to expand ~\eqref{eqn:MFm}
  to higher orders.
\bibitem{foot4} Homogeneous ordered phases were not reported in
  original studies of 1d flocking models---as far as we are aware---
  but we numerically checked that they are also present in the
  models of~\cite{Evans1999,Vicsek1d}.
\bibitem{foot12} Biological systems would typically corresponds to the
  limit $\eps=1$.
\end{thebibliography}
\end{document}